# Realization of BaZrS$_3$ chalcogenide perovskite thin films for optoelectronics

Xiucheng Wei[a], Haolei Hui[a], Chuan Zhao[a], Chenhua Deng[b], Mengjiao Han[c], Zhonghai Yu[d], Aaron Sheng[e], Pinku Roy[f], Aiping Chen[g], Junhao Lin[c], David F. Watson[e], Yi-Yang Sun[h], Tim Thomay[a], Sen Yang[d], Quanxi Jia[f], Shengbai Zhang[i], and Hao Zeng[a,*]

[a]Department of Physics, University at Buffalo, the State University of New York, Buffalo, NY 14260, USA

[b]Department of Chemistry, Taiyuan Normal University, Taiyuan 030619, China

[c]Department of Physics, Southern University of Science & Technology, Shenzhen 518055, China

[d]MOE Key Laboratory for Nonequilibrium Synthesis and Modulation of Condensed Matter, Xi'an Jiaotong University, Xi'an 710049, China

[e]Department of Chemistry, University at Buffalo, the State University of New York, Buffalo, NY 14260, USA

[f]Department of Materials Design and Innovation, University at Buffalo, the State University of New York, Buffalo, NY 14260, USA

[g]Center for Integrated Nanotechnologies, Materials Physics and Applications Division, Los Alamos National Laboratory, Los Alamos, NM 87545, USA

[h]State Key Laboratory of High Performance Ceramics and Superfine Microstructure, Shanghai Institute of Ceramics, Chinese Academy of Sciences, Shanghai 201899, China

[i]Department of Physics, Applied Physics & Astronomy, Rensselaer Polytechnic Institute, Troy, NY 12180, USA

* corresponding author: haozeng@buffalo.edu

**Abstract**

BaZrS$_3$ is a prototypical chalcogenide perovskite, an emerging class of unconventional semiconductor. Recent results on powder samples reveal that it is a material with a direct band gap of 1.7-1.8 eV, a very strong light-matter interaction, and a high chemical stability. Due to the lack of quality thin films, however, many fundamental properties of chalcogenide perovskites remain unknown, hindering their applications in optoelectronics. Here we report the fabrication of BaZrS$_3$ thin films, by sulfurization of oxide films deposited by pulsed laser deposition. We show that these films are n-type with carrier densities in the range of $10^{19}$-$10^{20}$ cm$^{-3}$. Depending on the processing temperature, the Hall mobility ranges from 2.1 to 13.7 cm$^2$/Vs. The absorption coefficient is > $10^5$ cm$^{-1}$ at photon energy > 1.97 eV. Temperature dependent conductivity measurements suggest shallow donor levels. By assuring that BaZrS$_3$ is a promising candidate, these results potentially unleash the family of chalcogenide perovskites for optoelectronics such as photodetectors, photovoltaics, and light emitting diodes.

**KEYWORDS**: chalcogenide perovskite, absorption coefficient, Hall effect, carrier mobility, defects

**Introduction**

Semiconductors can be considered as the backbone of modern society. They have found broad applications in computer chips, power electronics, optical sensors, light emitting diodes (LEDs), solid-state lasers, and solar cells. Most conventional semiconductors are covalent materials with four-fold coordination of both the cations and anions. During last decade, however, the organic-inorganic halide perovskites have attracted considerable attention, as they rival the conventional semiconductors for photovoltaics in an unprecedented way [1-8]. These are ionic materials, which are characterized by a higher coordination maximizing the Coulomb attraction between cations and anions. The strong ionicity is believed to minimize the possibility of forming deep level anti-site defects responsible for non-radiative carrier recombination. Compared to conventional semiconductors, the halide perovskites have unusually low carrier concentration (~ $10^{13}/cm^3$) and extremely long carrier lifetime (on the order of 1 µs) [9]. The power conversion efficiency of solar cells made of halide perovskites has witnessed a stellar rate of increase, from an initial PCE of 3.8% in 2009 [10] to above 25% in 2019 [11].

Specifically, perovskites refer to a class of crystalline compounds adopting the generic chemical formula $ABX_3$, where cation "B" has six nearest-neighbor anions. "X" and cation "A" sits in a cavity formed by eight corner-shared $BX_6$ octahedra. They demonstrate a rich spectrum of physical phenomena from 2D electron gas, ferroelectricity/piezoelectricity, ferromagnetism, colossal magnetoresistance, multiferroicity, ionic conductivity, to superconductivity [12-17]. The most commonly studied perovskites are the complex metal-oxides, where X is oxygen. Due to their multifunctionality and highly tunable physical properties, the oxides are an extremely important class of materials for technological applications.

Over the spectrum of covalency and iconicity, conventional semiconductors and oxide/halide perovskites are two extremes. Covalent bonding is directional, making the electronic and optical properties sensitive to bond distortions. In contrast, ionic bonding is often associated with a strong electron correlation, as the dielectric screening is reduced by the loss of shared valence electrons. Balancing ionicity with covalency therefore provides opportunities for discovering semiconductors with properties and performances unattainable in conventional semiconductors. In this regard, it is quite surprising that only limited effort has been devoted to the development of materials intermediate between covalency and ionicity.

Recently, chalcogenide perovskites have emerged as a novel class of semiconductor, where the anions are S and Se instead of O. They are more ionic than the conventional semiconductors but less so than either oxides or halides. Despite being synthesized more than a half century ago, these compounds have received little attention [18-22]. As a result, there is limited knowledge of their physical properties [23, 24]. The situation changed only recently, after we theoretically screened [25] 18 $ABX_3$ chalcogenide materials, predicting exciting semiconductor properties for photovoltaics. For example, several of them were found to be direct bandgap semiconductors combining both a strong light absorption and a good carrier mobility, which is a rare trait for semiconductors and are hence particularly attractive for optoelectronic applications. Subsequent experimental efforts have succeeded in synthesizing several of the prototypical chalcogenide perovskites as well as related phases including $BaZrS_3$, $CaZrS_3$, $SrTiS_3$, $SrZrS_3$[26], and $SrHfS_3$ [27, 28]. In particular, we further confirmed that $BaZrS_3$ possesses a distorted perovskite structure with a ~1.7 eV [27] bandgap and strong light absorption, in good agreement with our theory. The material was found to be exceptionally stable against pressure [29], moisture and oxidation [27].

The structural and physical properties show little degradation four years after the synthesis (see Fig. S1 in the Supporting Information, SI)

Besides, Niu et al. [26] synthesized and characterized $BaZrS_3$ and $SrZrS_3$. The latter has two different phases, with β-$SrZrS_3$ showing a bandgap of 2.13 eV and green light emission. They further demonstrated a giant optical anisotropy in hexagonal $BaTiS_3$ crystal [30]. Meng *et al.* [31] studied the $BaZr_{1-x}Ti_xS_3$ alloy system, suggesting a limited concentration range before phase separation takes place. Optical and thermoelectric properties of $SrHfSe_3$ and $Sr_{1-x}Sb_xHfSe_3$ were also studied [32]. Intense green luminescence characteristics were found in both undoped and heavily n- and p-doped $SrHfS_3$ [28]. On the theory front, Ruddlesden–Popper perovskite sulfides $A_3B_2S_7$ were proposed as a new family of ferroelectric photovoltaic materials in the visible [33]. Using machine learning, Agiorgousis *et al.* isolated $Ba_2AlNbS_6$, $Ba_2GaNbS_6$, $Ca_2GaNbS_6$, $Sr_2InNbS_6$, and $Ba_2SnHfS_6$, out of 450 chalcogenide double perovskites, as the most promising photovoltaic materials [34]. Guided by computational screening of ternary sulfides, Kuhar *et al.* also identified and synthesized $LaYS_3$ with a strong light absorption and photoluminescence as a promising candidate for photoelectrochemical water splitting [35].

While these studies reveal that chalcogenide perovskite and related compounds are indeed a unique family of optoelectronic materials with promises, a number of fundamental material properties such as the carrier type, concentration and mobility, optical absorption coefficient, and defect properties remain largely unexplored. This is due in large part to the lack of thin film samples, as most experiments have focused on powder or single crystal bulk samples. Lack of the thin film samples thus not only limits our basic understanding, but also becomes an obstacle against device applications.

In this paper, we report the first fabrication of BaZrS$_3$ thin films, by sulfurization of BaZrO$_3$ precursor films deposited by pulsed laser deposition. We show that films fabricated by this method are n-type with carrier density in the range of $10^{19}$-$10^{20}$ cm$^{-3}$. The Hall mobility ranges from 2.1 to 13.7 cm$^2$/Vs depending on processing temperature. The optical absorption coefficient is greater than $10^5$ cm$^{-1}$ at photon energy greater than 1.97 eV. Temperature dependent conductivity measurements suggest shallow donor levels. Although further optimization is needed, our present results suggest that BaZrS$_3$ thin films are promising for optoelectronic applications.

## Experimental

### Synthesis of BaZrS3 thin films

BaZrO$_3$ thin films with 100 nm thickness were deposited on sapphire substrates via a PLD-450 Pulsed Laser Deposition (PLD) system at 800°C. Oxygen partial pressure of 2.0 Pa was introduced into the deposition chamber with a background vacuum level higher than $10^{-5}$ Pa. The laser frequency and energy was set to be 5 Hz and 250 mJ, respectively. The as-deposited BaZrO$_3$ films were then loaded into a 2" quartz tube furnace for sulfurization. The sulfurization procedure lasted for 4 hours in an H$_2$/N$_2$ atmosphere at 900, 950, 1000, and 1050 °C, respectively. Sulfurization was also done at 1050 °C for 2 hours for the samples used for photodetector measurements. CS$_2$ was introduced at 800°C as the sulfur source, through H$_2$/N$_2$ gas bubbling at a flow rate of 20-25 standard cubic centimeters per minute (sccm).

A Rigaku Ultima IV X-ray diffraction (XRD) system with an operational X-ray tube power of 1.76 kW (40 kV, 44 mA) and Cu target source was used to acquire the X-ray diffraction pattern (XRD) for investigating the crystal structure. The XRD measurements were performed under theta/2 theta scanning mode and continuous scanning type with a step size of 0.02°. A Renishaw inVia Raman Microscope was used to measure the room temperature Raman and Photoluminescence (PL) spectra with a 1200 l/mm grating, 50× objective lens, and 514 nm laser. Time resolved PL (TRPL) on the BaZrS$_3$ film was done using an amplified ultrafast excitation laser (repetition rate 250 kHz) with a pulse duration of <200 fs and a wavelength of 400 nm. TRPL was collected with a microscope objective with an NA of 0.2 and spectrally and temporally detected on a Hamamatsu streak camera with a time-resolution of 32 ps. A Labsphere RSA-HP-8453 reflectance spectroscopy accessory attached to Agilent 8453 ultra-violet/visible (UV-vis) spectroscopy system was used to obtain the absorption spectrum. Surface morphology and energy-dispersive X-ray elemental analysis was performed using a Focused Ion Beam-Scanning Electron Microscope (FIB-SEM) – Carl Zeiss AURIGA CrossBeam with an Oxford EDS system.

**Atomic resolution imaging**

Atomically resolved high-angle annular dark field (HAADF) scanning transmission electron microscopy (STEM) images were acquired on a FEI Titan Themis Cube with an X-FEG electron gun and a DCOR aberration corrector operating at 300 kV. The semi convergence angle used was 30.1 mrad. The inner and outer collection angles for the STEM images (β1 and β2) were 54 and 143 mrad, respectively. The beam current was about 20 pA for the ADF imaging. All imagings were performed at room temperature. The EDX elemental mapping was acquired using the SuperEDX system equipped with four detector configurations.

**Device fabrication and optical and transport measurements.**

A two-terminal device was fabricated for photodetector measurements. Au electrodes were deposited using an e-beam evaporator through a mask with a thickness of 50 nm and a gap of 100 µm between the electrodes. A Keithley 2425 source/meter was used as a voltage source and a Keithley 6485 picoammeter was used as the current meter for I-V measurements under dark and illumination conditions. The illumination was provided by a diode laser at a wavelength of 532 nm with a power of 46 mW and spot size of 5 mm. For Hall effect measurements, the sample was cut into a Hall bar with dimension of 6 mm× 1 mm, and directly wired using silver paste. The transport data were acquired using a Quantum Design Physical Property Measurement System (PPMS) interfaced to a Keithley 2425 source/meter and two Keithley 2182 voltmeters. The schematic drawing of the devices are shown in **Figure S2** of the Supporting Information.

**Results and discussion**

**Structural characterizations:** The XRD pattern of a BaZrS$_3$ thin film sulfurized at 1050 °C is shown in Figure 1(a). It can be seen that the film is polycrystalline with no preferential orientation. The peaks can be matched to those of the standard file JCPDS 00-015-0327, showing that the sample possesses an orthorhombic distorted perovskite structure with Pnma space group. The extremely intense peak is the (0001) peak from the sapphire substrate. In Figure 1(b), the Raman spectrum of the BaZrS$_3$ film measured at room temperature is shown. Five broad peaks can be observed between 50-400 cm$^{-1}$, which are identified as $B_{1g}^1$, $A_g^4$, $B_{2g}^6$, $B_{1g}^4$, and $B_{1g}^5$ modes. The peak positions match with the theoretical predictions [27, 29] and published experimental reports [26, 27]. Figures 2(a) -2(d) show the SEM images of BaZrS$_3$ thin films sulfurized from 900 °C to 1050 °C for 4 hours, respectively. It can be seen clearly that the films are polycrystalline, with the grain sizes increasing with increasing sulfurization temperature. As shown in Fig. 2(e), the grain size increases from 0.43 ± 0.01 µm at 900 °C to 1.19 ± 0.06 µm at 1050 °C. Such grain sizes are in the optimal range for certain optoelectronic applications such as photovoltaics. A typical EDX spectrum of the BaZrS$_3$ film synthesized at 1050 °C is shown in Figure 2(f). The atomic ratio of Ba: Zr: S is found to be 1: 1.17: 2.91, close to the stoichiometric composition. The sulfur concentration ranges from 56.9 to 57.8% at 6 different locations measured on the same film, suggesting a small inhomogeneity accompanied by a slight sulfur deficiency (**Table S1** in Supporting Information). The sulfur deficiency is likely caused by sulfur vacancies [27], due to the high processing temperature. We notice that this is analogous to oxygen vacancies commonly observed in oxide perovskites [36-39]. At lower sulfurization temperatures, the S:Ba composition ratio decreases, as shown in **Fig. S5** in SI. However, this should not be interpreted as increasing sulfur vacancies, but instead incomplete conversion of BaZrO$_3$ [19]. Further investigations using aberration-corrected scanning transmission electron microscope (STEM) are shown in Fig. 2(g-i). The atomically resolved high-angle annular dark field (HAADF) image in Fig. 2(g) reveals a well-crystallized structure much resembles ABO$_3$ perovskites, consistent with the orthorhombic perovskite phase observed in the XRD measurements. Meanwhile, individual S, Zr, and Ba atomic column signals are clearly identified by the atom-by-atom EDX elemental mapping in another region of the crystal view along a different zone axis (Fig.2(h)), as shown in Fig. 2(i). These atomic resolution characterizations indicate that each grain maintains well crystalline structures, demonstrating the high quality of the as-synthesized BaZrS$_3$ polycrystalline thin film.

**Optical characterizations:**

The band gaps of BaZrS$_3$ with distorted perovskite structure have been theoretically calculated to be around 1.7-1.85 eV [25, 27, 31, 40-41] and experimentally verified to be within the same range [26, 27, 31]. In Figure 3(a), the UV-Vis absorption spectrum as a function of photon energy is plotted for the BaZrS$_3$ thin film synthesized at 1050 °C. The thin film geometry allows the extraction of the absorption coefficient α. It can be seen that α rises rapidly in the range of 1.7-1.8 eV, and exceeds $10^5$ cm$^{-1}$ at photon energy > 1.97 eV. This confirms the strong light absorption of the BaZrS$_3$. The band gap energy can be estimated from the Tauc plot in Fig. 3(b). The value is found to be 1.82 eV, slightly higher than our previously reported value for powder samples. The PL spectrum shows a broad peak centered at 1.81 eV, with a width of ~ 200 meV, in good agreement with the absorption measurement. The time resolved PL is shown in Fig. 3(c). A bi-exponential fitting [42, 43] is found to accurately describes the data. Two time constants extracted are $\tau_1$= 40 ns and $\tau_2$= 400 ns, which suggest that the sample may be inhomogeneous, *e.g.*, photo-carrier separation at domain boundaries could account for the slower recombination with a longer $\tau$.

**Transport measurements**

To quantify the characteristics of carrier transport, Hall measurements were performed on the four samples with sulfurization temperatures ranging from 900 to 1050 °C. All samples showed n-type conductivity, suggesting that the dominant carriers are electrons. This is likely due to the sulfur vacancies, as each of them will contribute to 2 excess electrons. The carrier density obtained from the Hall effect measurements ranges from $1.06\times10^{19}$ to $4.7\times10^{20}$ cm$^{-3}$, as shown in Fig 4(a), suggesting substantially high doping levels. The conductivity is in the range of 3.52 to 588 S/cm. Combing these results, the Hall mobility can be calculated and plotted as a function of sulfurization temperature in Fig. 4(c). It can be seen that the Hall mobility increases monotonically with increasing sulfurization temperature. This is expected as higher processing temperature leads to larger grain size and higher crystallinity, both suppresses carrier scattering. The mobility at 1050 °C is found to be 13.7 cm$^2$/Vs. This value is comparable to that of halide perovskites, such as MAPbI$_3$ [44]. The limiting factor seems to be the carrier concentration. We expect that with proper passivation and reduction of carrier density, this value can be improved by an order of magnitude.

Conductivity was measured as a function of temperature to elucidate the origin of the carriers. It can be seen that the conductivity increases with increasing temperature. Although the limited temperature and conductivity range makes differentiating transport mechanisms difficult, the conductivity vs. temperature can be best fitted by the Efros-Shklovskii variable range hopping model, where $G = G_0 \exp\left(-\sqrt{\frac{E}{k_B T}}\right)$ [45]. On the other hand, a fitting using Arrhenius Law shows a slightly larger deviation from experimental data (**Fig. S7** in SI). An activation energy is extracted to be 1.7 meV. This behavior may be understood as related to the average ionization energy of the donor levels to the conduction band edge, assuming a narrow distribution of density of states for such levels resulting from sulfur vacancies. Our results suggest that sulfur vacancies act as shallow defect levels that are readily ionized at room temperature. Further systematic studies are needed to identify and quantify the defects in BaZrS$_3$ thin films, and pinpoint the carrier transport mechanisms.

**Photodetector measurements**

To investigate the suitability of the BaZrS$_3$ films for optoelectronic applications, photodetector devices were fabricated using the samples sulfurized at 1050 °C. The I-V curves in the dark and under illumination are plotted in Fig. 5. For the sample sulfurized at 1050 °C for 4 hrs, the difference between the I-V curves in the dark and under illumination is small, due to the large carrier concentration of 2.7×10$^{20}$ cm$^{-3}$ (Fig. 5(a)). Postulating that prolonged high temperature processing leads to sulfur vacancy formation, increasing the carrier concentration, we reduced the sulfurization time in an attempt to enhance the photo-response. As can be seen from Fig. 5(b), reducing the sulfurization time to 2 hrs significantly increases the sample resistivity and decreases the dark current by three orders of magnitude, suggesting that carrier density due to sulfur vacancy is greatly reduced. We are not able to perform Hall effect measurement to determine the carrier concentration on this sample due to its high resistance. Nevertheless, as seen in Fig. 5(b), the photo-response is significantly enhanced, with an ON/OFF ratio of 20 at the bias voltage of 2 V. These results indicate that for better photodetector performance, it is crucial to suppress the dark current by further reducing the carrier concentration. We suggest that p-type doping (*e.g.* by Y or La) or sulfurization at high pressure can be considered to passivate the sulfur vacancy states.

## Conclusion

In conclusion, we have fabricated $BaZrS_3$ chalcogenide perovskite thin films using sulfurization of oxide films deposited by PLD. The films show exceptionally strong light absorption with an absorption coefficient $> 10^5$ cm$^{-1}$ at photon energy $> 2$ eV. The films are n-type with good carrier mobility of ~ 13.7 cm$^2$/Vs. The films are defect tolerant with shallow donors possibly originating from sulfur vacancies. Combined with its earth abundancy, high stability and non-toxicity, $BaZrS_3$ is a promising candidate for optoelectronics such as photodetectors, photovoltaics and light emitting diodes. As importantly, the findings here open the door for fabricating other high quality chalcogenide perovskite thin films for both fundamental studies and device applications.

## Acknowledgement:

Work supported by US NSF CBET-1510121, CBET-1510948, MRI-1229208, and DOE DE-EE0007364. Y. -Y.S. acknowledges support by NSFC under Grant 11774365.

**Figure captions**

**Figure 1.** (a) An X-ray diffraction pattern of a BaZrS$_3$ film sulfurized at 1050°C. The extremely intense peak comes from the (0001) peak of the sapphire substrate; (b) A Raman spectrum of the BaZrS$_3$ film measured at 300K.

**Figures 2.** (a)-(d) Typical SEM images of BaZrS$_3$ thin films sulfurized at temperatures of (a) 900, (b) 950, (c) 1000 and (d) 1050 °C, respectively. (e) The measured average grain size as a function of sulfurization temperature for BaZrS$_3$ thin films, obtained from corresponding SEM images. (f) An EDX spectrum of the BaZrS$_3$ film sulfurized at 1050 °C. The atomic ratio of Ba: Zr: S is found to be 1: 1.17: 2.91. Inset is an SEM image of the EDX measurement area. (g) Atomically resolved HAADF image of a BaZrS$_3$ thin film sulfurized at 900 °C. Inset is the Fast Fourier Transformation (FFT) of the image. (h) Atomically resolved HAADF image of another region of the sample viewed in a different zone axis. (i) EDX atom-by-atom elemental mapping corresponding to the center part of the area shown in (h).

**Figure 3.** (a) The UV-vis absorption spectrum; (b) Red curve: the Tauc plot derived from the absorption spectrum; Black curve: the PL spectrum; (c) Time resolved PL spectrum of the BaZrS$_3$ thin film sulfurized at 1050 °C for 4 hrs. Time constants τ$_1$ and τ$_2$ are found to be 40 ns and 400 ns, respectively from the bi-exponential fitting.

**Figure 4.** (a) Carrier density, (b) conductivity, and (c) Hall mobility of BaZrS$_3$ thin films as a function of sulfurization temperature for samples sulfurized for 4 hrs, obtained from the Hall effect measurements. (d) Conductivity plotted in logarithmic scale as a function of $T^{-1/2}$ for the BaZrS$_3$ film sulfurized at 1050 °C for 4 hrs.

**Figure 5.** The I-V curves of the photodetector devices measured in the dark (black curves) and under illumination (red curves) for the BaZrS$_3$ films sulfurized at (a) 1050 °C for 4 hrs, and (b) 1050 °C for 2 hrs.

# References

1. M. Liu, M. B. Johnston, H. J. Snaith, Efficient planar heterojunction perovskite solar cells by vapour deposition. Nature 501 (2013) 395-398.
2. J. Burschka, N. Pellet, S.-J. Moon, R. Humphry-Baker, P. Gao, M. K. Nazeeruddin, M. Gratzel, Sequential deposition as a route to high-performance perovskite-sensitized solar cells. Nature 499 (2013) 316-319.
3. G. Xing, N. Mathews, S. Sun, S. S. Lim, Y. M. Lam, M. Grtzel, S. Mhaisalkar, T. C. Sum, Long-Range Balanced Electronand Hole-Transport Lengths in Organic-Inorganic CH3NH3PbI3. Science 342 (2013) 344-347.
4. S. D. Stranks, G. E. Eperon, G. Grancini, C. Menelaou, M. J. P. Alcocer, T. Leijtens, L. M. Herz, A. Petrozza, H. J. Snaith, Electron-Hole Diffusion Lengths Exceeding 1 Micrometer in an Organometal Trihalide Perovskite Absorber. Science 342 (2013) 341-344.
5. N. J. Jeon, J. H. Noh, W. S. Yang, Y. C. Kim, S. Ryu, J. Seo, S. I. Seok, Compositional engineering of perovskite materials for high-performance solar cells. Nature 517 (2017) 476-480.
6. M. Gratzel, The light and shade of perovskite solar cells. Nat. Mater. 13 (2014) 838-842.
7. Q. Dong, Y. Fang, Y. Shao, P. Mulligan, J. Qiu, L. Cao, J. Huang, Electron-hole diffusion lengths > 175 μm in solution-grown CH3NH3PbI3 single crystals. Science 347 (2015) 967-970.
8. H. Zhou, Q. Chen, G. Li, S. Luo, T.-b. Song, H.-S. Duan, Z. Hong, J. You, Y. Liu, Y. Yang, Interface engineering of highly efficient perovskite solar cells. Science 345 (2014) 542-546.
9. D. W. de Quilettes, S. M. Vorpahl, S. D. Stranks, H. Nagaoka, G. E. Eperon, M. E. Ziffer, H. J. Snaith, D. S. Ginger, Impact of microstructure on local carrier lifetime in perovskite solar cells. Science 348 (2015) 683-686.
10. A. Kojima, K. Teshima, Y. Shirai, T. Miyasaka, Organometal Halide Perovskites as Visible-Light Sensitizers for Photovoltaic Cells. J. Am. Chem. Soc. 131 (2009) 6050-6051.
11. The NREL Research Cell Efficiency Records. http://www.nrel.gov/ncpv/, 2019 (accessed 11 October 2019).
12. T. Goto, T. Kimura, G. Lawes, A. P. Ramirez, Y. Tokura, Ferroelectricity and Giant Magnetocapacitance in Perovskite Rare-Earth Manganites. Phys. Rev. Lett. 92 (2004) 257201.
13. R. Ramesh, N. A. Spaldin, Multiferroics: progress and prospects in thin films. Nat Mater 6 (2007) 21-29.
14. A. A. Bokov, Z. G. Ye, Recent progress in relaxor ferroelectrics with perovskite structure, in: S. B. Lang, H. L. W. Chan (Eds.), Frontiers of Ferroelectricity: A Special Issue of the Journal of Materials Science, Springer US, Boston, 2007, pp. 31-52.
15. K. F. Wang, J. M. Liu, Z. F. Ren, Multiferroicity: the coupling between magnetic and polarization orders. Adv. Phys. 58 (2009) 321-448.
16. D. I. Khomskii, Multiferroics: Different ways to combine magnetism and ferroelectricity. Journal of Magnetism and Magnetic Materials 306 (2006) 1-8.
17. B. Jeroen van den, I. K. Daniel, Multiferroicity due to charge ordering. J. Phys. Condens. Matter 20 (2008) 434217.
18. H. Hahn, U. Mutschke, Untersuchungen über ternäre Chalkogenide. XI. Versuche zur Darstellung von Thioperowskiten. Z. Anorg. Allg. Chem. 288 (1957) 269-278.
19. A. Clearfield, The synthesis and crystal structures of some alkaline earth titanium and zirconium sulfides. Acta Crystallogr. 16 (1963) 135-142.
20. R. Lelieveld, D. J. W. Ijdo, Sulphides with the GdFeO3 Structure. Acta Crystallogr. B36 (1980) 2223-2226.
21. Y. Wang, N. Sato, T. Fujino, Synthesis of BaZrS3 by short time reaction at lower temperatures. J. Alloy. Comp. 327, (2001) 104-112.
22. C.-S. Lee, K. M. Kleinke, H. Kleinke, Synthesis, Structure, and electronic and physical properties of the two SrZrS3 modifications. Solid State Sci. 7 (2005) 1049-1054.

23. M. Ishii, M. Saeki, Raman and Infrared Spectra of BaTiS3 and BaNbS3. Phys. Stat. Sol. (b) 170 (1992) K49.
24. M. Ishii, M. Saeki, M. Sekita, Vibrational spectra of barium-zirconium sulfides. Mater. Res. Bull. 28 (1993) 493-500.
25. Y.-Y. Sun, M. L. Agiorgousis, P. Zhang, S. Zhang, Chalcogenide perovskites for photovoltaics. Nano Lett. 15 (2015) 581-585.
26. S. Niu et al., Bandgap Control via Structural and Chemical Tuning of Transition Metal Perovskite Chalcogenides. Adv. Mater. 29 (2017) 1604733.
27. S. Perera, H. Hui, C. Zhao, H. Xue, F. Sun, C. Deng, N. Gross, C. Milleville, X. Xu, D. F. Watson, B. Weinstein, Y.-Y. Sun, S. Zhang, H. Zeng, Chalcogenide perovskites – an emerging class of ionic semiconductors. Nano Energy 22 (2016) 129-135.
28. K. Hanzawa, S. Iimura, H. Hiramatsu, H. Hosono, Material Design of Green-Light-Emitting Semiconductors: Perovskite-Type Sulfide SrHfS3. J. Am. Chem. Soc. 141 (2019) 5343-5349.
29. N. Gross, Y.-Y. Sun, S. Perera, H. Hui, X. Wei, S. Zhang, H. Zeng, B. Weinstein, Stability and Band-Gap Tuning of the Chalcogenide Perovskite BaZrS3 in Raman and Optical Investigations at High Pressures. Physical Review Applied 8 (2017) 044014.
30. S. Niu, G. Joe, H. Zhao, Y. Zhou, T. Orvis, H. Huyan, J. Salman, K. Mahalingam, B. Urwin, J. Wu, Y. Liu, T. E. Tiwald, S. B. Cronin, B. M. Howe, M. Mecklenburg, R. Haiges, D. J. Singh, H. Wang, M. A. Kats, J. Ravichandran, Giant optical anisotropy in a quasi-one-dimensional crystal. Nat. Photonics 12 (2018) 392-396.
31. W. Meng, B. Saparov, F. Hong, J. Wang, D. B. Mitzi, Y. Yan, Alloying and defect control within chalcogenide perovskites for optimized photovoltaic application. Chemistry of Materials 28 (2016) 821-829.
32. N. A. Moroz et al., Insights on the Synthesis, Crystal and Electronic Structures, and Optical and Thermoelectric Properties of Sr1–xSbxHfSe3 Orthorhombic Perovskite. Inorg. Chem. 57 (2018) 7402-7411.
33. H. Wang, G. Gou, J. Li, Ruddlesden–Popper perovskite sulfides A3B2S7: A new family of ferroelectric photovoltaic materials for the visible spectrum. Nano Energy 22 (2016) 507-513.
34. M. L. Agiorgousis, Y.-Y. Sun, D.-H. Choe, D. West, S. Zhang, Machine Learning Augmented Discovery of Chalcogenide Double Perovskites for Photovoltaics. Adv. Theory Simul. 2 (2019) 1800173.
35. K. Kuhar, A. Crovetto, M. Pandey, K. S. Thygesen, B. Seger, P. C. Vesborg, O. Hansen, I. Chorkendorff, K. W. Jacobsen, Sulfide perovskites for solar energy conversion applications: computational screening and synthesis of the selected compound LaYS3. Energy Environ. Sci. 10 (2017) 2579-2593.
36. J. Scott, M. Dawber, Oxygen-vacancy ordering as a fatigue mechanism in perovskite ferroelectrics. Appl. Phys. Lett. 76 (2000) 3801-3803.
37. C. Park, D. J. Chadi, Microscopic study of oxygen-vacancy defects in ferroelectric perovskites. Phys. Rev. B 57 (1998) R13961.
38. K. R. Poeppelmeier, M. Leonowicz, J. Longo, CaMnO2.5 and Ca2MnO3.5: New oxygen-defect perovskite-type oxides. J. Solid State Chem. 44 (1982) 89-98.
39. C. Michel, L. Er-Rakho, B. Raveau, The oxygen defect perovskite BaLa4Cu5O13.4, a metallic conductor. Mater. Res. Bull. 20 (1985) 667-671.
40. J. W. Bennett, I. Grinberg, A. M. Rappe, Effect of substituting of S for O: The sulfide perovskite BaZrS3 investigated with density functional theory. Phys. Rev. B 79 (2009) 235115.
41. J. M. Polfus, T. Norby, R. Bredesen, Protons in oxysulfides, oxysulfates, and sulfides: a first-principles study of La2O2S, La2O2SO4, SrZrS3, and BaZrS3. J. Phys. Chem. C 119 (2015) 23875-23882.
42. S. A. Kumar, S. Prasanth, C. Joseph, A novel UV-emitting poly (vinylidene fluoride-hexafluoropropylene)-CQD composite material for optoelectronic applications. AIP Conf Proc, 2082 (2019) 030001.

43. M. Joy, E. Anabha, S. Gopi, B. Mathew, S. Kumar, A. Mathews, Structural and optical profile of a multifunctionalized 2-pyridone derivative in a crystal engineering perspective. Acta Cryst. C74 (2018) 807-815.
44. L. M. Herz, Charge-carrier mobilities in metal halide perovskites: fundamental mechanisms and limits. ACS Energy Lett. 2 (2017) 1539-1548.
45. A. L. Efros, B. I. Shklovskii, Coulomb gap and low temperature conductivity of disordered systems. J. Phys. C: Solid State Phys. 8 (1975) L49.

**Figures**

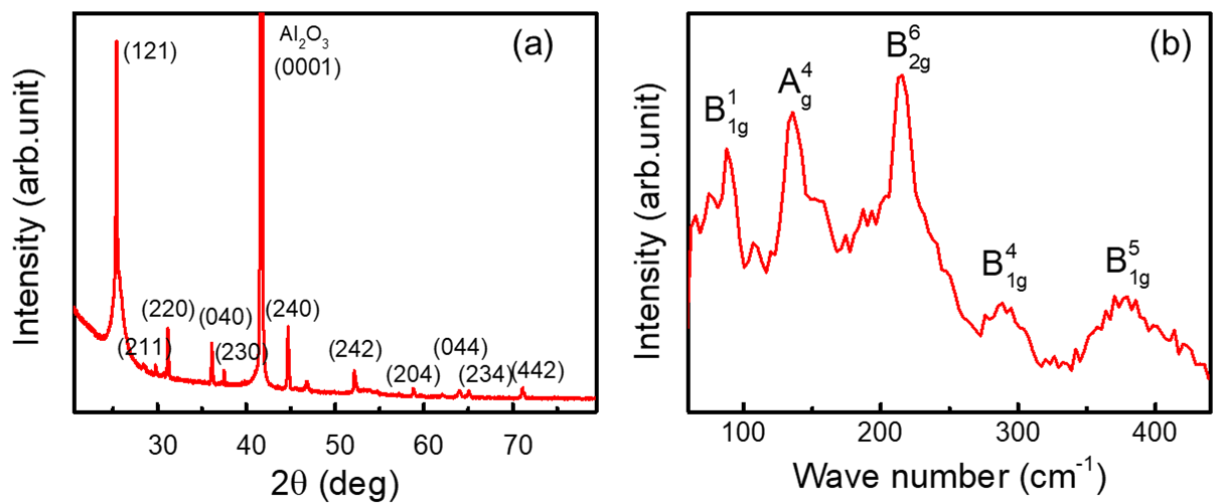

Figure 1
17

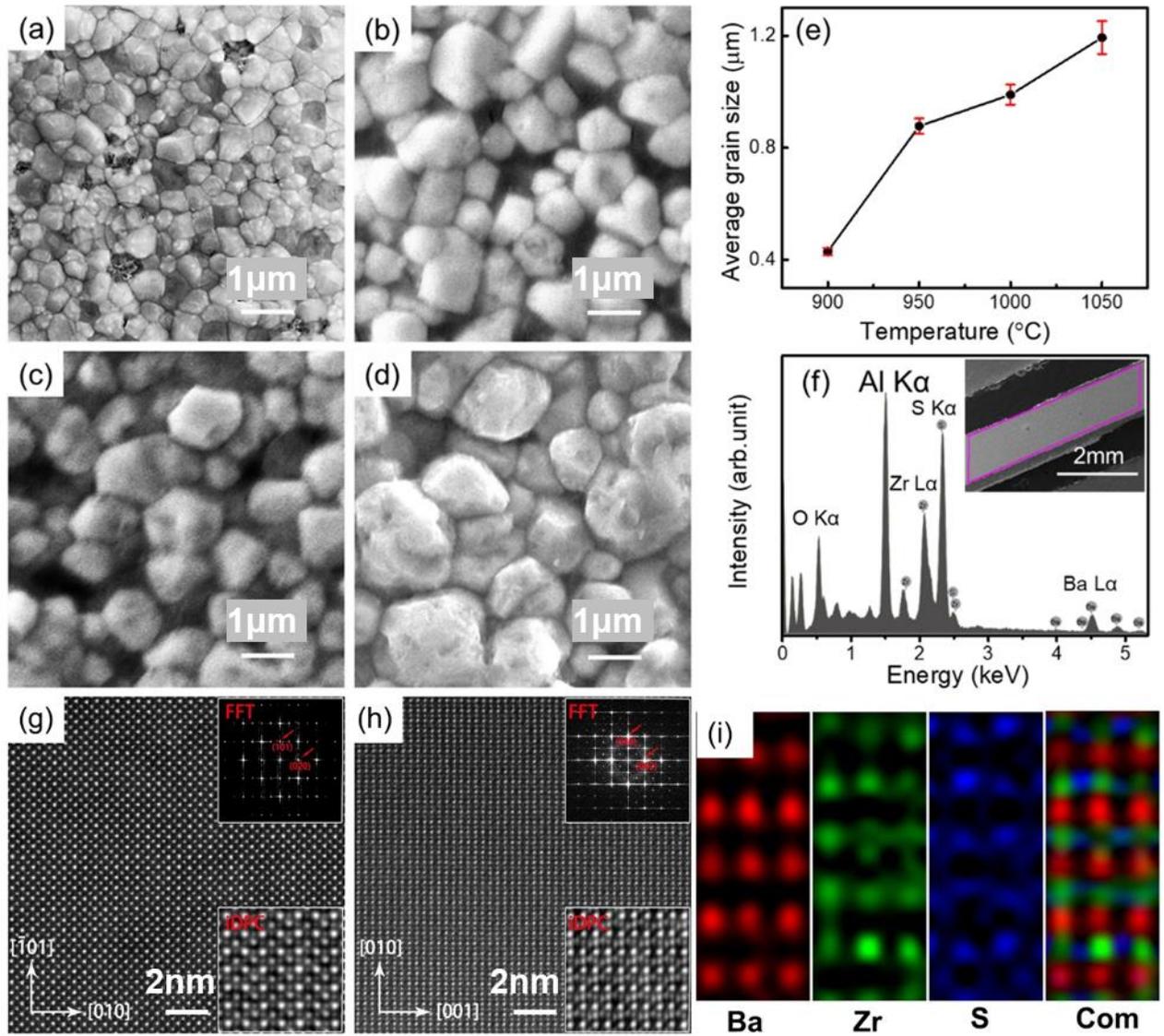

**Figure 2**

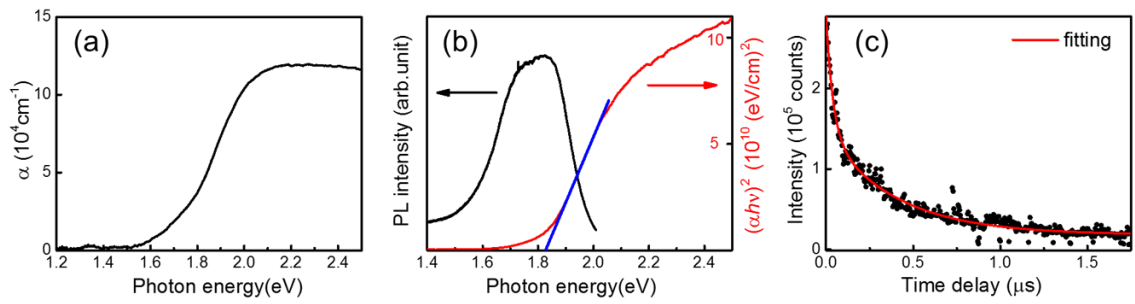

**Figure 3**

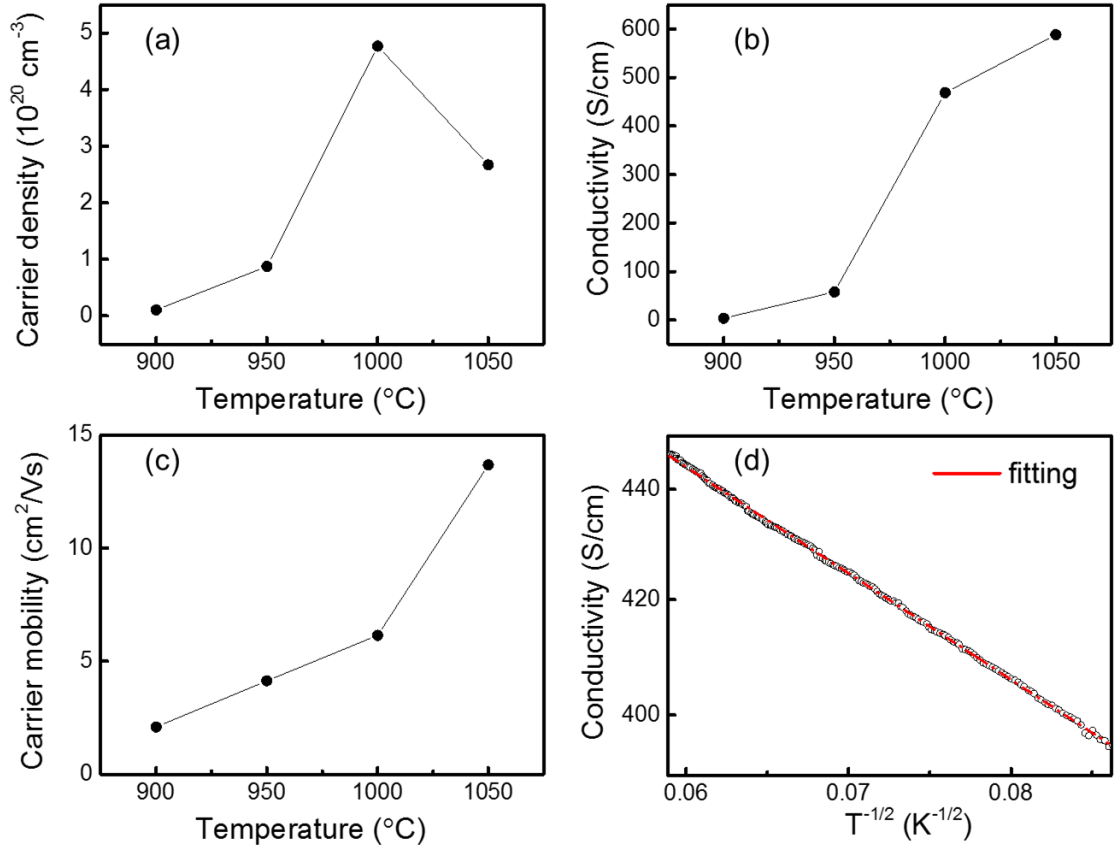

**Figure 4**

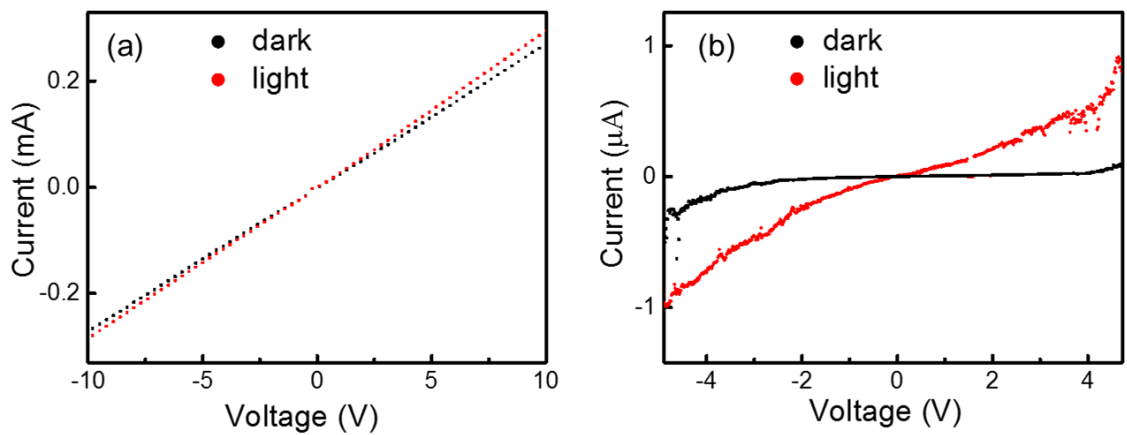

**Figure 5**

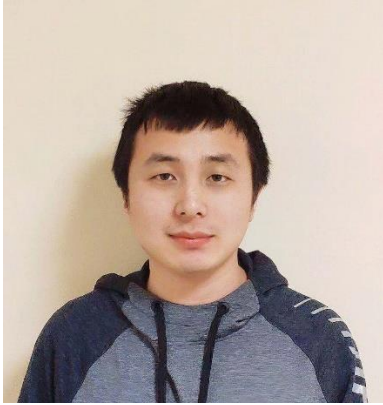

**Xiucheng Wei** received his B.S. degree in physics from Xi'an Jiaotong University, China. He is currently a graduate student in the department of physics at University at Buffalo. His research interests focus mainly on novel photovoltaic semiconducting materials and devices.

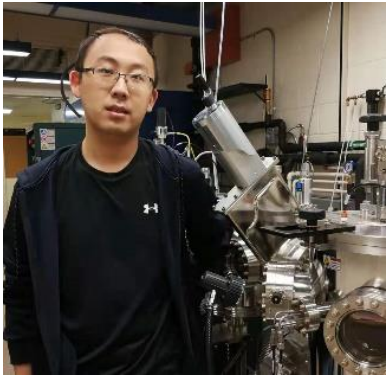

**Chuan Zhao** received his PhD degree in Physics from University at Buffalo in 2019. He is currently working at Intel Corporation, Dalian, China. His main research field is about 2D materials such as single-layer $WSe_2$ and $WS_2$.

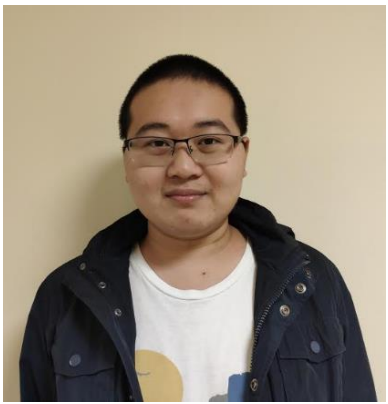

**Haolei Hui** received his bachelor's degree in Xi'an Jiaotong University in 2018. He is currently a PhD student in department of physics at University at Buffalo. He is focusing on novel chalcogenide perovskite as photovoltaic materials.

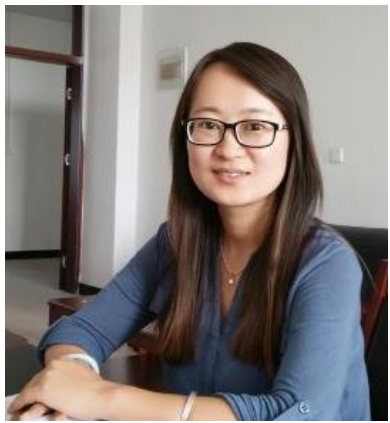

**Chenhua Deng** received her Ph.D. degree from Shanxi Normal University in 2016. She is currently an associate professor at Taiyuan Normal University. Her current research interest includes semiconductor spintronic, magnetic recording media, chemistry and physics of nanoscale materials.

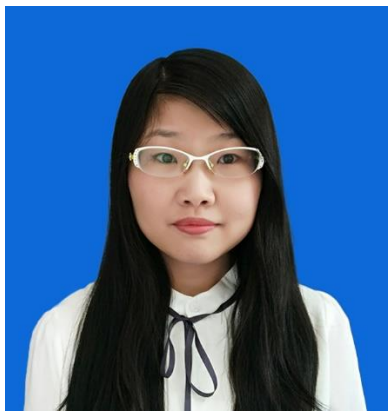

**Mengjiao Han** received her Ph.D. in the field of materials science and engineering from Institute of Metal Research, University of Chinese Academy of Sciences in 2019. Now she works as a postdoc at the Southern University of Science and Technology and focuses on the correlations between the atomic structure and the property of two-dimension (2D) materials.

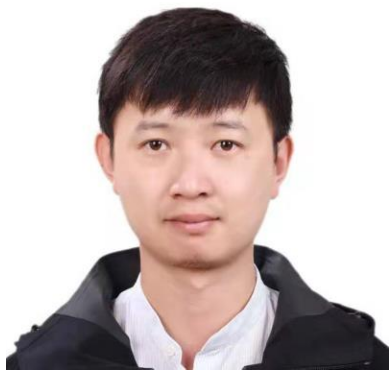

**Zhonghai Yu** received his bachelor of science degree from Taiyuan University of Technology in 2014. He is currently a doctoral student working on his PhD, focusing on Chalcogenide Perovskites solar cell at Xi'an Jiaotong University since 2015. He will be a visiting PhD student at the University at Buffalo.

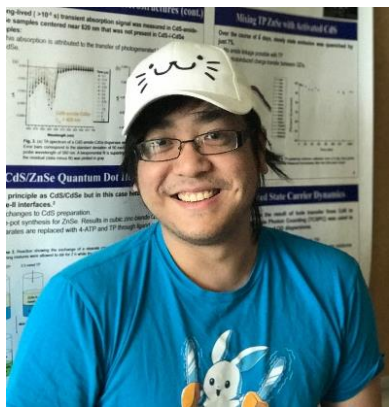

**Aaron Sheng** received his B.S in chemistry from University at Buffalo in 2015. He is currently a 5[th] year student in the Department of Chemistry at University at Buffalo in the Watson Lab. His research involves synthesis and characterizing of excited-state charge-transfer in vanadium oxide/quantum dot heterostructures.

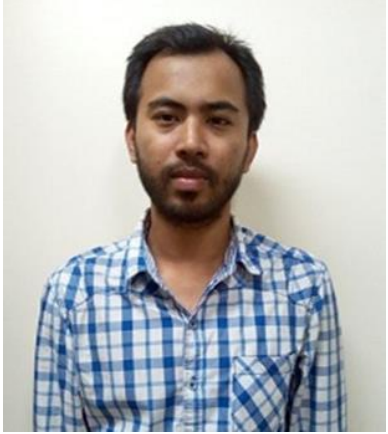

**Pinku Roy** received his MTech. in Materials Science and Engineering from IIT Kanpur, India. He is currently a graduate student in Materials Design and Innovation department in University at Buffalo. His research focuses on X-ray diffraction, epitaxial thin film, electronic and magnetic materials.

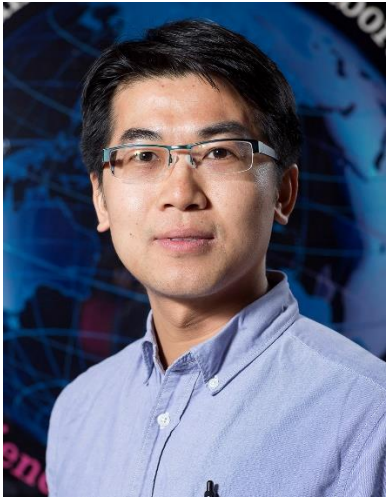

**Aiping Chen** is currently a staff scientist in Center for Integrated Nanotechnologies, one of five Nanoscale Science Research Centers funded by U.S. DOE Office of Science. His research focuses on synthesizing quantum oxide heterostructures, tuning functionalities via microstructure, defect, strain and interface engineering, and exploring their applications in microelectronics.

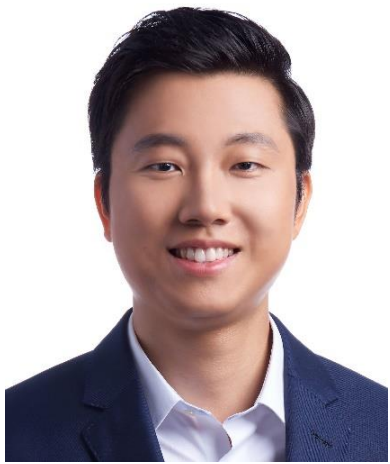

**Junhao Lin** got his PhD in Physics from Vanderbilt University, USA, in 2015, and joined Southern University of Science and Technology (SuSTech) as an associated professor after 3 years postdoctoral training in National Advanced Institute of Science and Technology (AIST), Japan. His research interest includes analysis of complex defect structures in novel layered materials, real time in-situ observation of the dynamical processes in structural transition of materials under various environmental stimulations, and the phonon behavior of 2D materials as probed by monochromatic valence electron energy loss spectroscopy (VEELS).

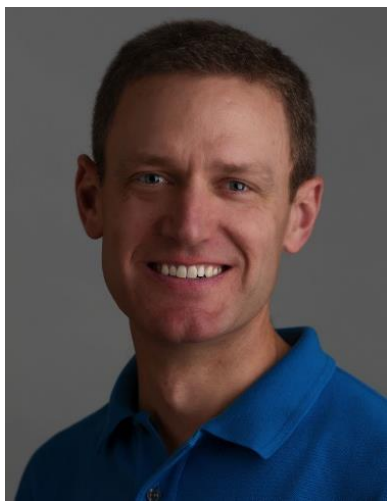

**David Watson** received his Ph.D. in chemistry from Princeton University in 2001. He joined the University at Buffalo in 2004 after a postdoc at Johns Hopkins University. He is currently a professor and chair of the Department of Chemistry. His research involves synthesizing nanostructured materials and interfaces and characterizing their excited-state charge-transfer reactivity.

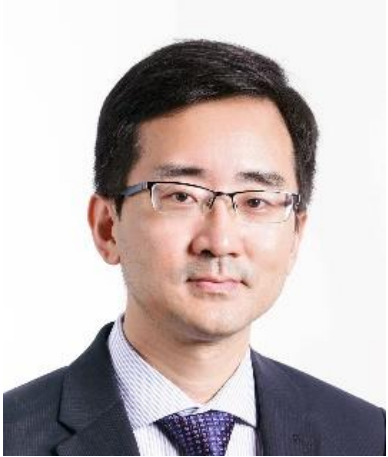

**Yi-Yang Sun** received his Ph.D. in physics from National University of Singapore (NUS) in 2004. Since then, he has worked as a postdoc at NUS, National Renewable Energy Laboratory and Rensselaer Polytechnic Institute (RPI), USA. In 2010, he was appointed Research Assistant Professor and later Research Scientist at RPI. In 2017, he assumed a Professor position at Shanghai Institute of Ceramics, Chinese Academy of Sciences. He has been working on first-principles study of energy-related materials.

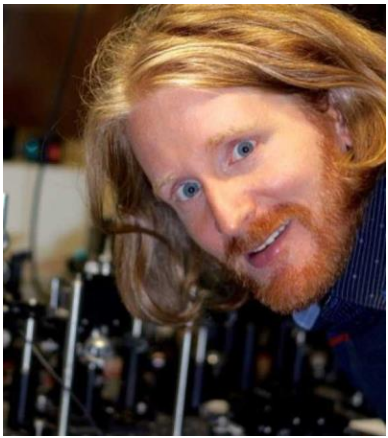

**Tim Thomay** received his Dr. rer. nat. in Physics from the University of Konstanz, Germany in 2009. He joined the Department of Physics at the University at Buffalo in 2019 after a 4-year postdoc at NIST Gaithersburg, MD and after joining the EE department at Buffalo as a researcher. His research interests are in the ultrafast dynamics of low dimensional solid-state structures and devices and low photon number quantum optics.

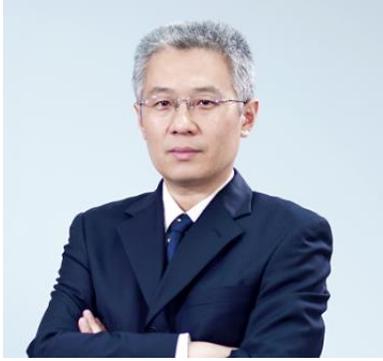

**Sen Yang** received his Ph.D. in materials physics from the Xi'an Jiaotong University (XJTU), China in 2005. He joined the National Institute for Materials Science, Japan in 2005 as a JSPS (Japan Society for the Promotion of Science) post-doctor. In the year of 2010, he came back to XJTU and was promoted to full professor in 2013. His research interests are in magnetism and magnetic materials, smart materials, phase transition and so on.

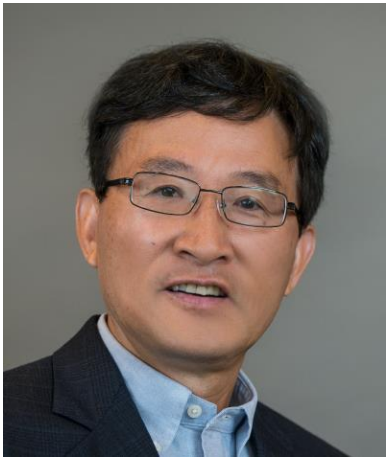

**Quanxi Jia** is an Empire Innovation Professor and National Grid Professor of Materials Research at the University at Buffalo (UB). Prior to joining UB in 2016, he had worked at Los Alamos National Laboratory for 24 years, with the last two years serving as the co-Director and then Director of the Center for Integrated Nanotechnologies, a US Department of Energy Nanoscale Science Research Center operated jointly by Los Alamos and Sandia National Laboratories. His research focuses on nanostructured and multifunctional materials, with a particular effort on the synthesis and study of processing-structure–property relationships of epitaxial films for energy and electronic applications.

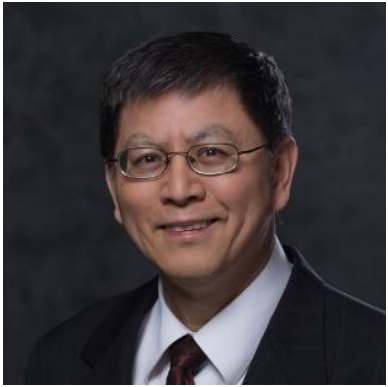

**Shengbai Zhang** received his Ph. D. in physics from University of California at Berkeley in 1989. He moved to Xerox PARC as a postdoc, before joining the National Renewable Energy Laboratory in 1991. In 2008, he became the Senior Kodosky Constellation Chair and Professor in Physics at Rensselaer Polytechnic Institute. His expertise is first-principles theory, modeling, and calculation. Recent work involves chalcogenide perovskites for photovoltaic, ultrafast phase change memory materials, topological carbon networks, non-equilibrium growth, excited state dynamics, and unconventional two-dimensional materials and excitonic insulators. He is a Fellow of the American Physical Society since 2001.

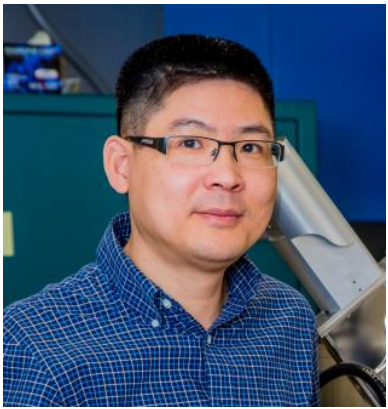

**Hao Zeng** received his Ph.D. in physics from the University of Nebraska-Lincoln in 2001. He joined the Department of Physics at the University at Buffalo in 2004 after a 3-year postdoc at IBM T.J. Watson Research Center, and was promoted to full professor in 2014. His research interests are in nanoscale magnetism and magnetic materials, spintronics, 2D materials and unconventional semiconductor materials and devices.